\documentclass[%
 reprint,
 amsmath,amssymb,
 aps,
]{revtex4-1}

\usepackage{amsmath}
\usepackage{graphicx}
\newcommand{\veps}{\varepsilon}
\newcommand{\om}{\omega}
\newcommand{\gam}{\gamma}
\newcommand{\bee}{\begin{equation}}
\newcommand{\ene}{\end{equation}}
\graphicspath{{pics/}}
\usepackage{wrapfig}

\begin{document}

\preprint{APS/123-QED}

\title{Disorder-induced Purcell enhancement in  nanoparticle chains}


\author{Mihail Petrov}
\altaffiliation[Also at ]{the Institute of Photonics, University of Eastern Finland, Yliopistokatu 7,
80100 , Joensuu, Finland}
\affiliation{Center of Nanophotonics and Metamaterials, ITMO University,  Birjevaja line V.O. 14, 199034, Saint-Petersburg, Russia}
\email{trisha.petrov@gmail.com}


\date{\today}

\begin{abstract}
In this paper we report on  numerical study of plasmonic nanoparticle chains with long-range dipole-dipole interaction.  We have shown that  introduction of positional disorder gives a  peak  in the density of resonant states (DOS) at the frequency of individual nanoparticle resonance. This peak is  referred to Dyson singularity in one-dimensional disordered structures  [Dyson F., 1953] and, according to our calculations, governs the spectral properties of local DOS. This provides disorder-induced Purcell enhancement that can found its applications in random lasers and for SERS spectroscopy. We stress that this effect relates not only to plasmonic nanoparticles but to an arbitrary chain of nanoparticles or atoms with resonant polarizabilities.

\end{abstract}



\maketitle

\section{Introduction}


Plasmonic and dielectric nanoparticle chains have been actively studied due to their subwalength waveguiding properties in a number of papers for the last decade \cite{Wei2004,Simovski2005,Citritn2005,Alu2011,Campione2011, Noskov2012}. Rapidly developing self-assembly fabrication methods allow production of tens nanometers scale structures with simple and cost-effective techniques \cite{Esteban2012, Kitching2013, Solis2012, Chervinskii2013,Klinkova2014}. One of the main features of self-assembly methods is the randomness and disorder in fabricated  structures.   Spatial order and periodicity play key role in the process of efficient energy transport, and introduction of disorder leads to suppressing the transmission efficiency \cite{Alu2010,Markel2007,Ruting2011}. However,  the role of  disorder in photonics and plasmonics have been recently reconsidered. The experimental advances in  random lasing \cite{Cao2005,Wiersma2008} stimulated studies on light transport in disorder media \cite{Wiersma2013} and photon management in strongly scattering media \cite{Segev2013,Vynck2012}. In this paper we discuss the  utilization  of disorder to induce Purcell enhancement in  resonant chains.    

Since the early works of Mott and Anderson  one-dimensional (1D)  disordered structures have been attracting intensive interest of the researchers. Worth noting is the monography of Gredeskul, Pastur, and Pitaevski \cite{Lifshits1988} almost fully dedicated to disorder in 1D.  In photonics disordered one-dimensional photonic crystal  have been studied in a number of papers \cite{Liew2010, Poddubny2012,Greshnov2008}. We focus  on  nanoparticle chains that are quasi one-dimensional as they are embedded in 3D, but the chain excitations propagate along  the chain  direction only. In this prospective   it was shown \cite{Alu2010,Markel2007} that  introduction of disorder in nanoparticle chains  stimulates  scattering and increase losses. On the other hand, the randomness in plasmonic structures can be beneficial and  give rise into giant fluctuations of local fields and to accumulation of energy in ``hot spots'' \cite{Genov2005,Stockman2001} that finds its application in SERS. \par

 In this paper we report on how  disorder can be utilized to control and engineer optical properties of resonant nanoparticle chain. We demonstrate that in one-dimensional chain the positional disorder stimulates formation of special modes that gives its contribution to  DOS. Such behavior was predicted by F. Dyson in 1953 \cite{Dyson1953} for a chains of mechanical oscillators with random values of spring stiffness. He showed that at zero energy there exists divergence in DOS function. The divergent DOS is related   to phonon spectra of solids \cite{Parshin1998,Beltukov2011},   
excitonic structures in disordered 1D J-aggregates \cite{Fidder1991,Kozlov1998,Avgin1999}, and  mathematically  all these systems can be merged within the theory of random matrices \cite{Wigner1955, Mehta2004}. Being disorder induced and, in this sense, disorder protected, the effect of divergent DOS can be implemented  for  local density of states (LDOS) control and  spontaneous emission engineering via Purcell effect \cite{Purcell1946,Sapienza2011,Aigouy2014}.   This can open a route for fabrication of random lasers in one-dimensional structures and for additional SERS enhancement on disorder plasmonic chains.   

The manuscript is structured as follows: in section I we  formulate our approach to the problem considering disordered nanoparticle chain as an array of radiating plasmonic dipoles.  In section II  the problem is treated  in quasi statical (QS)  and  nearest neighbor (NN) interaction limits. We show that there exists a singularity in the DOS function at the frequency of an isolated nanoparticle resonance. The considered simple model  partially explains the basic physics lying beyond the  discussed effects. In section III more realistic  approach with long-range interaction and retardation effects is proposed. We include near, intermediate, and far fields into consideration and show that the Dyson peculiarity still persists in the DOS spectra.   Finally, in the section IV we will discuss the Dyson peak in LDOS and demonstrate how it is influenced by losses. In the Appendix \ref{AppendixA} we describe the  method of eigen frequency calculation. In Appendix \ref{AppendixB} The case of non-1D (planar) nanoparticle array is considered  on the example of double-line chain. The sufficient difference of DOS spectrum comparing to 1D chain is discussed. 

\section{Formulation of the problem}         

	We consider a chain of plasmonic nanoparticles as an example of coupled dipoles with Drude like dielectric permittivity $\veps(\om) =\veps_{\infty}-{\om_{p}^{2}}/{\om(\om+i\gam)}$. Here $\omega_{p}$ is the plasma frequency of metal and $\gamma$ is the damping constant.  We consider spherical nanoparticles in vacuum with  polarizability 
$$
\alpha=R^{3}\dfrac{\veps(\omega) -1}{\veps(\omega) +2},    
$$
where $R$ is the nanoparticle radius. Neglecting  losses and assuming  $\veps_{\infty}=1$  the resonant frequency  of an  individual nanoparticle has the form $\omega_{0}=\omega_{p}/\sqrt{\veps_{\infty}+2}$.  The  polarizability can be written as 

$$\alpha(\om)=R^{3}\dfrac{\om_{0}^{2}}{\om_{0}^{2}-\om^{2}}$$
%
 To  account on retardation one can use substitution \cite{Novotny2012}
\begin{equation}
\label{retard}
\qquad \dfrac{1}{\alpha}\rightarrow \dfrac{1}{\alpha}-\dfrac 23 i k^{3},
\end{equation}
where $k$ is the  wave vector. 
\begin{figure}
\includegraphics[scale=0.7]{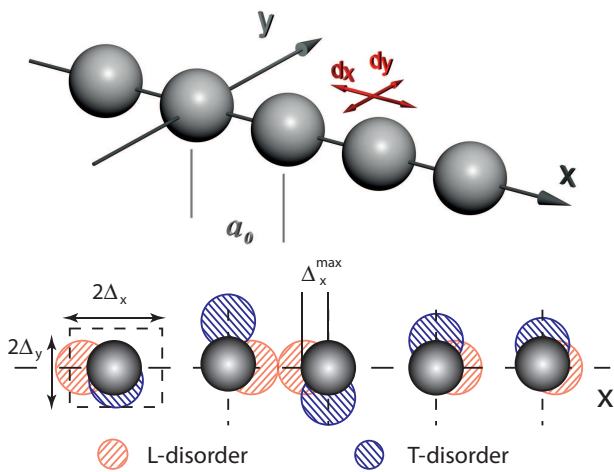}
\caption{Regular nanoparticle chain with fixed distance $a_{0}$ between nanoparticles (top). The types of  the spatial disorder (bottom): longitudinal (L) disorder relates to random position shift in $x$ direction;   transverse (T) disorder relates to random position shift in $y$ direction. $\Delta^{max}$ is  the  maximal range of a shift that provides nonoverlaping condition.      }
\label{Disorder}
\end{figure}
 Dipole interaction of particles  splits  the resonant frequency $\om_{0}$ and takes off the polarization degeneracy. For infinite periodic chain the set of eigen frequencies forms bands for transverse (T) and longitudinal (L) polarization \cite{Weber2004,Simovski2005,Campione2011}. In order to study the effects of disorder we will consider finite chains consisting of $N$  nanoparticles. We can write down the system of equations describing the dipole moments of nanoparticle 
\bee
\label{Gen1}
\vec{d_i}=\alpha(\omega)\sum_{j=1,\ j\neq i}^{N}\tensor{G}_{ij}(\omega)\vec{d_{j}}, \qquad i=1..N.
\ene
 Here $\tensor{G}_{ij} (\omega)=\tensor{G}(|\vec{r}_{i}-\vec{r}_{j}|,\omega)$ is dyadic Green function \cite{Novotny2012} that represents the field of  a point dipole placed at the coordinate $\vec{r}_{j}$ and calculated at the point $\vec{r}_{j}$. The homogeneous problem describing the system has the form \cite{Weber2004,Dong2009,Deng2011} 
 \bee
\label{Gen2}
{\bf H(\omega)}{ \bf d}=\dfrac 1 {\alpha(\omega)} \bf{d}
\ene
where {\bf H} is $3N\times3N$  block-matrix, at  the $ij$-th $(1\le i,j\le $N$)$ position there is matrix representation of $\tensor{G}(\omega)_{ij}$ tensor, and {\bf d} is a block-vector of length $3N$  containing components of  $\vec{d_{i}}$ at i-th position. The homogeneous system (\ref{Gen2}) has non-trivial solution if the condition
\bee
\det\left({\bf H(\omega)}-\dfrac1{\alpha(\omega)}{\bf I}\right)=0
\label{Gen3}
\ene
is fulfilled, where ${\bf I}$ is the unity matrix. Solving this equation one can find all the resonant frequencies of the system \cite{Weber2004}. However, for  ensembles with large $N$ this appears to be a complicated problem. We propose using perturbation method and  represent $\bf{H}(\om)=\bf{H}(\om_{0})+\delta\bf{H(\om_{0})}$ basing on weak dependence of matrix {\bf H} on $\om$. In this case the inverse polarizability plays the role of $\bf{H(\om_{0})}$ matrix eigen number \cite{Markel2006,Dong2009}. The eigen frequencies can be found within the first order precision: $\omega_{n}=\omega_{n}^{0}+\omega_{n}^{1}$. More detailed description of this approach is described  in the Appendix \ref{AppendixA}. 
\begin{figure*}
\includegraphics[scale=0.6]{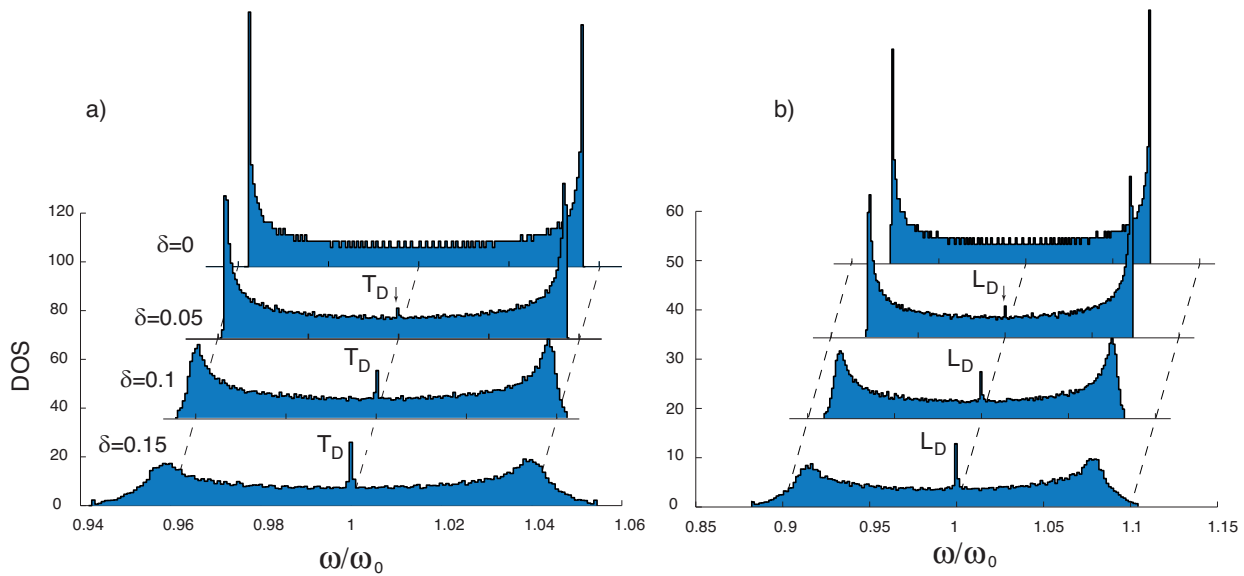}
\caption{DOS in the QS limit with amplitude of L-disorder. a) T-polarized modes b) L-polarized modes. N=800. The results are calculated for $a_{0}=3R$, N=800, and averaged over 50 realizations.}
\label{DOS_QS}
\end{figure*}

\subsection{Introduction of disorder}

In this paper we consider a plasmonic chain with randomly varying   nanoparticle positions that is also referred as non-diagonal disorder. Each nanoparticle can be randomly shifted around its position in the regular chain. We limit the consideration with  $x-y$ plane perturbation only, keeping the chain to be planar (see Fig. \ref{Disorder}). Such geometry is the most prospective from the point of possible  technological realization: nanoparticle ensembles are commonly fabricated on the top of planar substrates \cite{Solis2012,Slaughter2012,Chervinskii2013}. Each nanoparticle center can be varied for a random value $dx$ or $dy$. We use uniform distribution of the i-th nanoparticle shift $dx_{i}(dy_{i})$ in the range $-\Delta_{x(y)}\le dx(dy)_{i}\le \Delta_{x(y)}$. We introduce the relative parameter of disorder $\delta_{x(y)}=\Delta_{x(y)}/a_{0}$, which  is  limited with its maximal value $\delta_{max}(a_{0})$ to satisfy the condition of non-overlaping. For instance, the value of $\delta_{x}^{max}\approx0.18$ for $a_{0}=3R$.  

We define two types of disorder as it is shown in Fig.~\ref{Disorder}: (a) $\Delta_{x}\neq0, \ \Delta_{y}=0$ longitudinal disorder (L-disorder) and  (b) $\Delta_{x}=0, \ \Delta_{y}\neq 0$ transverse disorder (T-disorder). We need to stress that L-disorder  keeps  splitting of T and L polarizations, meanwhile T-disorder mixes T and L  polarizations. We also will discuss the case of TL disorder that is superposition of both types of disorder. We  will pay much attention to  L-disordered chains as they have simpler description but possess all the key properties of the considered systems.

\section{Simple model: QS limit and nearest-neighbor interaction}

We start with the consideration of  nanoparticle size and interparticle distance small   comparatively to the  wavelength (QS limit). The matrix {\bf H} depends on frequency through the term $\om r/c$ that goes to zero in the QS limit. 
The matrix {\bf H} becomes frequency independent and, thus, real symmetric. We come to exact eigen value problem with inverse polarizabilities as eigen values $\lambda_{k}$. The eigen frequency for each eigen value $\lambda_{k}$ can be found from the relation $1/\alpha(\om_{k})=\lambda_{k}$.      
$$
{\bf H}(0){\bf d}=\lambda_{k}{\bf d}=\dfrac{1}{\alpha_{k}}{\bf d}
$$

In the simplest case of NN interaction the matrix {\bf H} has the  form: 

\begin{eqnarray}
\label{Hmatrix}
\begin{array}{cl}
$${\bf H}_{i,i}=0$$& \\
$${\bf H}_{i+1,i}={\bf H}_{i,i+1}=\dfrac{-1 (2)}{a_{i,i+1}^{3}}$$\  &\mbox{for T(L)-polarization}
\end{array}
\end{eqnarray}

For regular  chain the distance $a_{i,i+1}$ between the particles is constant $a_{i,i+1}=a_{0}$ and the  dispersion relation $\om(q)$  has the form \cite{Weber2004}: 
\begin{eqnarray}
w_{T(L)}^{2}(q)=&w_{0}^{2}\left(1+g_{T(L)}\cos(qa)\right),\nonumber\\
g_{T}=2\dfrac{R^{3}}{a_{0^{3}}}  &g_{L}=-4\dfrac{R^{3}}{a_{0^{3}}} \nonumber
\end{eqnarray}
where $q$ is quasi vector. 

 The DOS function then can be described by a simple formula that exhibits  van Hove singularities at the band edges: 
\begin{align}
DOS_{T(L)}(\om)=\frac{2\om}{a_{0}\sqrt{\om_{0}^{4}g_{T(L)}^{2}-\left(\om^{2}-\om_{0}^{2}\right)^{2}}} \nonumber
\end{align}

In the case of   L-disordered  chain $a_{i,i+1}=a_{0}+\Delta_{i+1}-\Delta_{i}$  is a random number  with mean value equal to $a_{0}$ and $\Delta_{i}$ are the nanoparticle shifts in $x$ direction and are given by uniform probability distribution. Thus, the system is described with two-diagonal symmetric matrix filled  with random numbers. Such matrix was considered in the classical paper by F. Dyson \cite{Dyson1953}. One of the results demonstrated in this paper was a singularity of DOS at the zero eigen values in the limit $N\rightarrow \infty$ for non diagonal disorder that is the case of our study. In terms of polarizability the condition of $\lambda=0$ is satisfied at the  frequency $\om_{D}=\om_{0}$ at which inverse polarizability has root  $1/\alpha(\om)\sim1-\om^{2}/\om_{0}^{2}$.  In the Fig. \ref{DOS_QS} the DOS of nanoparticle chain is presented in the QS limit and NN interaction for different value of disorder parameter $\delta_{x}$. Similar behaviors were already obtained in \cite{Fidder1991,Kozlov1998,Avgin1999} for excitons in one-dimensional disorder structures. The Frenkel hamiltonian in tight binding limit considered in these papers  has the form of  (\ref{Hmatrix}).  

The  QS nearest-neighbor interaction limit gives the logarithmic divergence of the DOS function in the vicinity of zero eigen value  \cite{Dyson1953}. However, since Dyson classical paper  there are various discussions on the nature of the divergent states.  In \cite{Theodorou1976} it was shown that the  states are extended, and, thus, delocalized. The later studies by  \cite{Kozlov1998} dispute this claim. Our calculations of participation ratio also demonstrate that the localization length is finite, however the \cite{Avgin1999} reports that this may be due to  finite size effect. The appearance of this singularity can be understood within the  perturbation theory. Kozlov et.al. \cite{Kozlov1998} showed that in QS limit introducing disorder as perturbation gives zero   shift   of energy levels in the middle of the band  that results in increasing of  DOS.  

Since the early  works of Dyson and Wigner active development of random matrix theory has started, and sparse random matrices play an important role among them.  In several papers  \cite{Evangelou1992,Rogers2008,Beltukov2011} it has been shown that   the singularity of DOS for  zero eigen value is  tightly related to sparsity of random matrices. The constructed matrix (\ref{Hmatrix}) is  just the case of a sparse random matrix. The random matrix theory predicts that this effect can be observed in the vicinity of zero eigen value in a wide class of random matrices. In our terms, zero value eigen values correspond to resonance in polarizability. Consequently, divergent DOS behavior will be observed in quite general class  of interacting resonant oscillators and plasmonic particle chain is just an example. 

\begin{figure}[htbp]
\begin{center}
\includegraphics[scale=0.7]{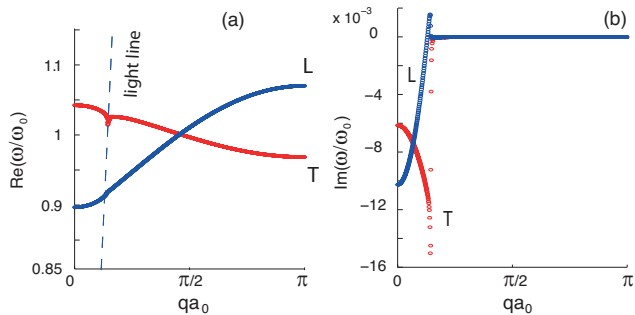}
\caption{(a) Dispersion relation of regular nanoparticle chain with retardation for T and L modes. The dispersion of light in free space (light line) is also shown. (b) Dependence of imaginary parts of eigen frequencies on quasi vector for T and L modes. The results are calculated for  $a_{0}=3R$, $\gamma=0$, $\kappa=0.15$, N=400.   }
\label{DispRetard}
\end{center}
\end{figure}

\section{Retardation and long-range interaction}
The previous studies by \cite{Fidder1991,Kozlov1998} considered long-range dipole-dipole interaction only in QS regime, i.e. the interaction was governed by near fields that decrease as $\sim 1/r^{3}$.  Considering plasmonic chain the long range radiation effects play important role.  We considered fully retarded model with intermediate  ($\sim 1/r^{2}$) and far  ($\sim 1/r$) fields that are the major interaction terms at long distances. We demonstrate that despite stronger long-range coupling one still can observe the peculiarity at the band center.
\begin{figure}
\begin{center}
\includegraphics[scale=1]{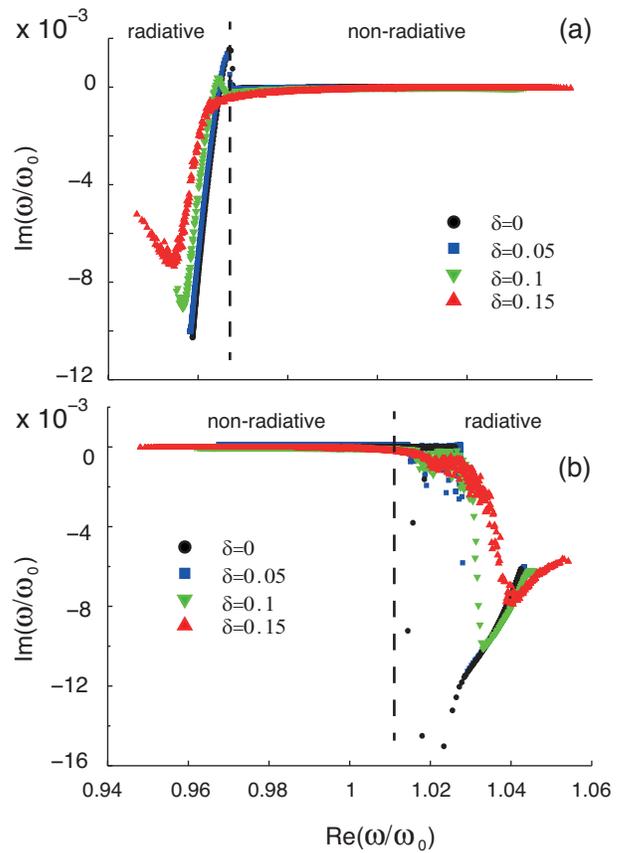}
\caption{The imaginary parts of eigen frequencies for (a) L-polarization and (b) T-polarization for different magnitude of L-disorder. The parameters of calculation are taken the same as in the Fig. \ref{DispRetard}. Results are obtained after averaging over 50 realizations.    }
\label{DispRetardDrd}
\end{center}
\end{figure}

To include the retardation effects we consider frequency dependent matrix ${\bf H (\om)}$ and  use substitution (\ref{retard}) to account on dipole emission that  introduces  losses in the system. We neglect Drude losses $\gamma=0$ at this stage, but  will add them in the last section of the manuscript. To find the eigen frequency of the system (\ref{Gen2}) we apply approximate method described in the Appendix \ref{AppendixA} section. The main parameter that defines the dependence of {\bf H} matrix on frequency is $\kappa=\om_{0}R/c$. Indeed, the eigen frequencies will lie around  $\om_{0}$ and parameter $\kappa$ shows the retardation strength, i.e. typical phase shift at the scales of nanoparticle radius. In this paper we are limited with small values of $\kappa$ as approximate method of calculation based on perturbation theory  diverges for large $\kappa$, thus, for all the calculations below we take $\kappa=0.15$. 
\begin{figure*}[htbp]
\includegraphics[scale=0.7]{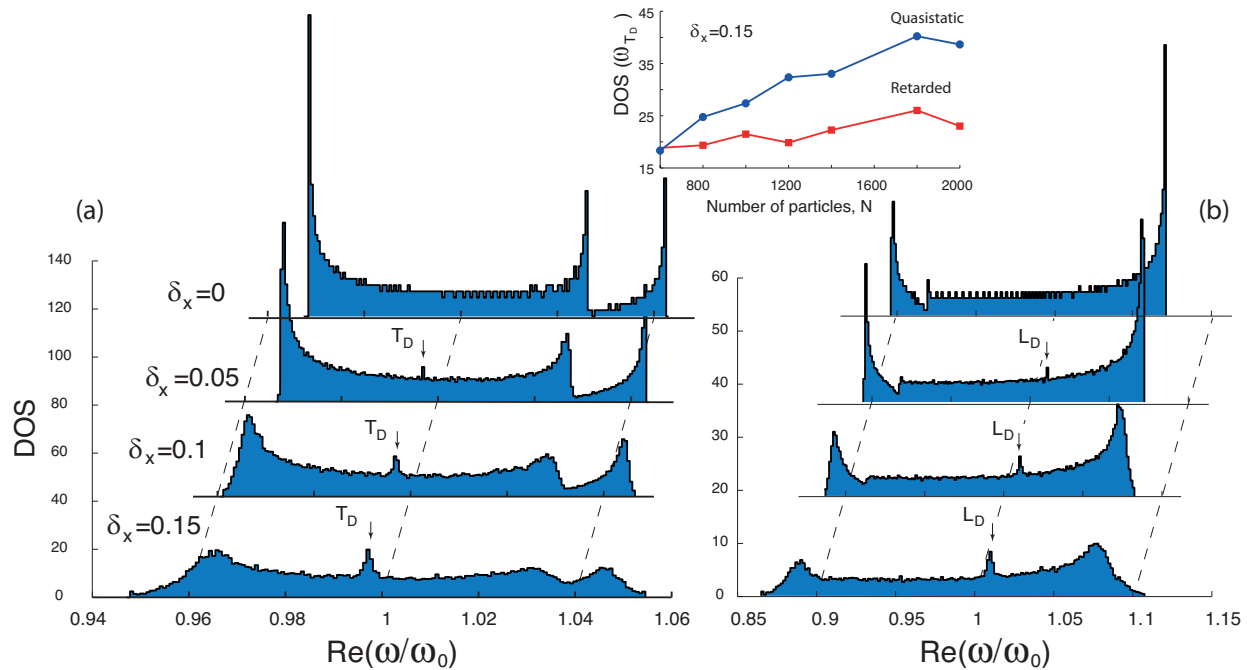}
\caption{DOS for L-disordered chain with retardation for T-polarized modes (a) and L-polarized modes (b). The results are calculated for $a_{0}=3R$, $\kappa=0.15$, $\gamma=0$, $N=800$ and averaged over 50 realizations. Inset:  DOS calculated at the frequency $\om_{T_{D}}$ corresponding to $T_{D}$ peak in quasi state limit (see Fig. \ref{DOS_QS})  (circles) and with retardation and long-range interaction (squares) depending on chain length $N$ at fixed magnitude of disorder $\delta_{x}=0.15$.   }
\label{DosRetard}
\end{figure*}

\subsection{Regular chain}

To understand the influence of disorder with account on retardation we again start with the regular chain. The periodical nanoparticle chains with account on retardation and losses have been extensively studied during last decades \cite{Weber2004,Campione2011,Markel2007} because of their waveguiding properties and application in plasmonics. The real and imaginary parts of regular chain eigen frequencies are shown in Figure \ref{DispRetard} (a,b)  for T and  L modes. The dispersion relation depicted in Fig. \ref{DispRetard} (a) was obtained by sorting the eigenmodes in the  descending order of node numbers in the eigen vector. The Bloch vector corresponds to number of zeroes as:
$$
q=\dfrac{\pi (n+1)}{Na_{0}}
$$

 According to \cite{Weber2004} the T-modes lying above the light line are well  coupled to light and are highly radiative (superradiative)  that is confirmed by the large negative value of Im$(\om_{i})$ .  On the contrary, the modes below the light line are non-radiative (waveguiding regime) and are almost decoupled with light however the imaginary part is nonzero either. The strong hybridization of waveguiding  modes with light is seen in the vicinity of the light line crossing with the dispersion curve.  The calculation method shows discrepancy for L-modes close to $ka\sim 0.2$ where the imaginary part becomes positive.    

 To demonstrate how disorder affects the eigen spectrum we plotted  imaginary parts of eigen frequencies in the Fig.\ref{DispRetardDrd} for T-modes (a) and for L-modes (b). We see that the disordered structures inherit the properties of ordered systems for both T- and L-modes and we again can divide the spectral region into highly radiating and non-radiating regions.  Destruction of periodicity increases losses in non-radiative region due to additional scattering of Bloch waves on defects, but  in the radiating region losses are decreased because disorder   suppresses the superradiative  regime.

To describe the properties of disordered structures we again introduce DOS similar to the QS case. However, in the case of retardation the eigen frequencies are complex, and we will plot the DOS depending on real part of eigen frequency Re$(\om)$  (see Fig. \ref{DosRetard}). 

The DOS function for regular chain has complicated structure due to the strong hybridization of chain modes with light. With the increase of disordered the DOS function becomes more and more homogeneous. But despite the long-range interaction and retardation one can  see the divergence of DOS function near the frequency $\om_{0}$ similarly to the case of NN limit. However,  the DOS peak is now shifted for T and L-modes comparatively to the case shown in Fig. \ref{DOS_QS} due to long-range interaction effects. The shift of the  peak   corresponds to the shift of  band central frequency \cite{Kozlov1998} and this shift has opposite sign for T and L modes. The  DOS peak intensity at  increases with the increase of chain length $N$ (see Fig. \ref{DosRetard} inset) both in QS and retarded case which reflects it divergent character. 

\section{Planar (TL) disorder }
\label{TLdrd}
\begin{figure}
\begin{center}
\includegraphics[scale=0.7]{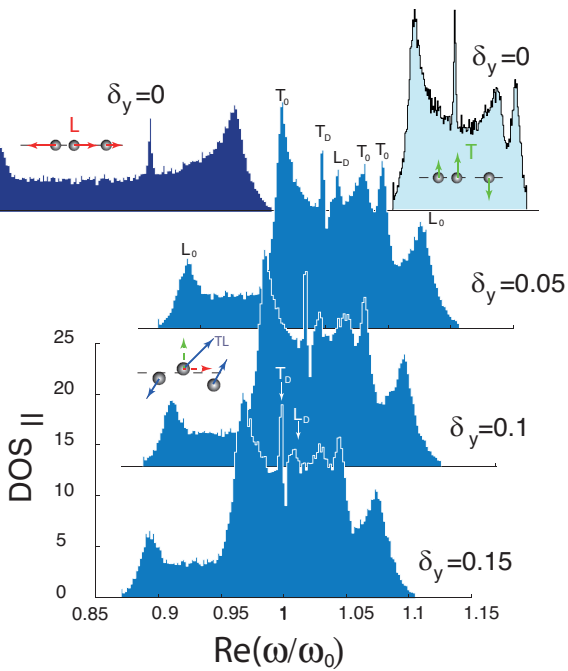}
\caption{DOS function for   TL-disordered chain with $\delta_{x}=0.15$ and different values of $\delta_{y}$. The DOS functions of T and L polarized states for purely L-disordered ($\delta_{y}=0$) chain are shown separately at the top of the figure.   Each dipole moment in TL-disordered mode has $x$ and $y$  components (see inset). For weak disorder one can divide modes into mainly T- and mainly L-polarized. The peaks related to DOS of ordered chain are denoted as $T_{0}$ and $L_{0}$ along with Dyson peaks $T_{D}$ and $L_{D}$. The parameters of calculations are the same as in the Fig. \ref{DosRetard}.}
\label{TL_DOS}
\end{center}
\end{figure}

Till now we have considered only L-disordered that conserves  splitting of modes into T and L polarizations. Introduction of  TL-disorder partially  mixes mode polarization. The  polarization of eigen modes lying in the   $x$-$y$ plane will be mixed, and the $z$-component of vector will be still independent and behave similarly to $y$-polarized modes in the case of L-disorder. Thus, we will not discuss $z$-polarized modes in this section. To emphasize that we will refer to DOS$_{||}$ meaning the density of resonant states with polarization vector lying in the $x$-$y$ plane.  The influence of  TL-disorder on DOS is depicted in the Fig.~\ref{TL_DOS}. One can see that introduction of week disorder  $\delta_{y}=0.05$ mixes the polarization  not significantly and the DOS function is roughly the sum of T (light blue) and L-polarization (dark blue) DOS functions in purely L-disordered chain. We mark the peaks originating from the ordered T and L polarizations as T$_{0}$ and L$_{0}$ respectively. The Dyson peaks are marked as T$_{D}$ and L$_{D}$. One can see that the increase of T-disorder affects the L$_{D}$ peak mainly and for $\delta_{y}=0.15$ the L$_{D}$ is  almost smeared out. The T$_{D}$ peak on the contrary does not change significantly. Here we can conclude that the transversal Dyson mode ($T_{D}$) is stable with  respect to transversal disorder in contrary with longitudinal mode.

In the Appendix \ref{AppendixB} we have also considered the special case of doubli-line chain with TL disorder to demonstrate that the features of Dyson singularity originates from   1D character of interaction. Considering two parallel chains of nanoparticle dramatically changes the DOS function and the fine structure around $\om_{0}$ disappears.   
\section{Local DOS and Purcell enhancement}

We have demonstrated that the disorder in nanoparticle chain leads to peculiarity in general DOS function near the band center, which relates to Dyson singularity of 1D  disordered system.  In nanophotonics the role of LDOS function is even of more importance. The LDOS  is connected to DOS via simple relation $D(\om)=\int \rho({ \bf r},\om) d^{3} {\bf r}$, thus, one can expect that LDOS  properties posses similar spectral features as DOS. 

\begin{figure}
\begin{center}
\includegraphics[scale=0.6]{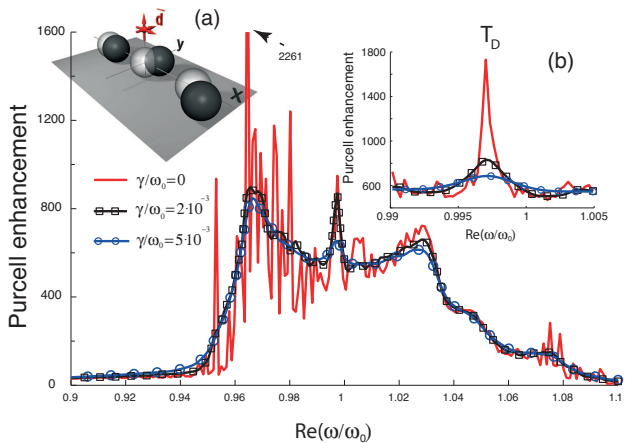}
\caption{Spectrum of Purcell enhancement factor calculated at  the point $(0,0,2R)$ and  plotted for different values of losses in metal. Radiation losses are also accounted. \emph{Insea:} a) the  point of Purcell enhancment calculation; b) zoomed  Purcell enhancement curve around Dyson singularity. The  number of particles  in chain  $N=101$. The parameters of calculation: $\delta_{x}=0.15,\ \delta_{y}=0.15,\ \kappa=0.15$. The $\gamma=0$ curve was obtained after averaging over 2000 realizations, and the other curves  were averaged over 500 realizations. }
\label{LDOS}
\end{center}
\end{figure}

 To obtain the LDOS function we have calculated the dyadic Green function $\widehat{G}(\bf{r},\bf{r},\om)$ of disordered chain \cite{Novotny2012}. We have placed the electric dipole source at the position above the chain, as it is shown in Fig.\ref{LDOS}, at the center of periodic chain at the coordinate $(0,0,2R)$,  and  calculated   chain  response to the excitation by the source  in  dipole approximation. The $x,y,\mbox{\ and\ } z$ components of electric field at the point of $x,y,\mbox{\ and\ } z$-oriented dipole source correspondingly  give us the  diagonal elements of the dyadic tensor, and  the LDOS can be found as:
$$
\rho({\bf r},\om)=\rho_{0}+\frac{2\om}{\pi c^{2}}\mbox{Im}{\left(\mbox{Tr}\ \widehat{G}(\bf{r},\bf{r},\om)\right)}
$$

\begin{figure}
\begin{center}
\includegraphics[scale=0.7]{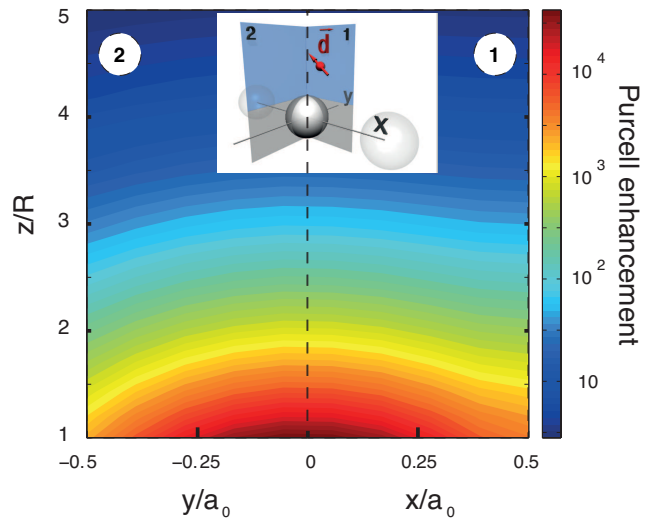}
\caption{LDOS enhancement distribution  for TL disordered chain at the frequency of Dyson singularity pointed  by arrow in the fig. \ref{LDOSxy}. Due to the symmetry of the problem LDOS is plotted only in a half of the unit cell as shown in the \emph{inset} (planes 1 and 2) for $z\ge R$.  The  number of particles  in chain  $N=101$. The parameters of calculation: $\delta_{x}=0.15,\ \ \delta_{y}=0.15,\ \ \kappa=0.15,\ \ \gamma/\om_{0}=2\cdot10^{-3}$. The result was obtained after  averaging over  500 realizations. }
\label{LDOSxy}
\end{center}
\end{figure}
We normalize  LDOS over $\rho_{0}=\om^{2}/\pi^{2}c^{3}$ that is LDOS  in vacuum that gives us the Purcell enhancement (PE) factor \cite{Purcell1946} averaged over dipole orientations. The calculated LDOS is depicted in Fig.\ref{LDOS} for TL disordered chain.  The spectrum lines are plotted for different values of losses in metal. For $\gamma=0$ only retardation losses are accounted. Low losses lead to strong fluctuations of the PE due to very narrow resonances. The fluctuations are particularly strong for the frequency range  below the $\om_{0}$ that corresponds to non-radiative regime of T-modes for periodic chain where radiative losses are small (see Fig.\ref{DispRetard} (a)). Adding Drude losses we significantly suppress the fluctuations and achieve smooth LDOS line already  for several hundreds of iterations.  One can see that the spectrum of PE replicates the DOS spectrum of T-modes  (see fig.\ref{DosRetard} (b)) with clear Dyson peak. It is relatively weak as the length of the chain was taken  $N=101$  to reduce computation time (DOS function presented in Fig.~\ref{DosRetard} was computed for $N=800$). Absence of  $L_{D}$ peak relates to weak coupling of dipole source with L-modes. Indeed,  coupling to L modes occurs only via near fields only for a dipole polarized in $x$ direction. The coupling via far field is suppressed as there are no $x$-components of electric field. 

   The spatial distribution of PE depending on dipole source position is plotted in the Fig.~\ref{LDOSxy} for $z>R$ at the frequency corresponding to ${T_{D}}$. Recalling that LDOS should be periodic in $x$ direction, we plot PE distribution  over one unit cell only.  The calculated PE is symmetric in the unit cell in  the $x-y$ plane respectively to   nanoparticle position in regular chain. Thus, we plot only half of the unit cell crossection in $x-z$ and $y-z$ planes as sown in Fig.~\ref{LDOSxy} inset. One can see that the PE distribution rapidly decrease away from the chain. We need to mention that the dipole model should not be valid close to nanoparticle surface around $z\lesssim R$ and $x\approx0,\ y\approx0$. 
   
   We again would like to stress that  the calculated LDOS corresponds to long range-interacted and retarded  system. The presence of peculiarity close to nanoparticle resonance $\omega_{0}$ is related to existence of resonant behavior of polarizability (existence of a root in the inverse polarizability $1/\alpha$). Thus, we predict the similar LDOS properties for a  general class of resonantly interacting oscillatory chain of different nature.

\section{Conclusion}

We have studied resonant nanoparticle chain with positional (non-diagonal) disorder. We have shown that disorder induces the divergence  in DOS function in the vicinity of individual nanoparticle resonance  that is associated with Dyson singularity. This divergence is stronger in the quasi statical, nearest neighbor  interaction limit, but also was found in long-range interacting system with retardation effects. Such DOS structure is an internal feature of one-dimensional systems and can not be observed in 2D and 3D  arrays. We have shown that positional disorder affects T and  L modes differently that is related to the character of dipole-dipole coupling. The LDOS function  inherits the structure of DOS and has a peak   in the vicinity of an  individual nanoparticle resonance.   This results may be applied for the observation of Purcell enhancement and for the development of spontaneous time emission engineering in general case of interacting oscillators with resonant polarizabilities. 
We believe that this effect can be applied for random lasing in one-dimensional structures and for additional enhancement of SERS signal in the vicinity of disordered chains.   
\section{Acknowledgement}

 This work was financially supported by Government of Russian Federation, Grant 074-U01, and by Academy of Finland, Mobility Grant. The author would like to thank  Yuri Kivshar  for useful comments on the manuscript; fruitful scientific discussions with Pavel Belov, Andrey Bogdanov, and Ivan Iorsh have made this publication possible.    

\begin{figure}
\begin{center}
\includegraphics[scale=0.5]{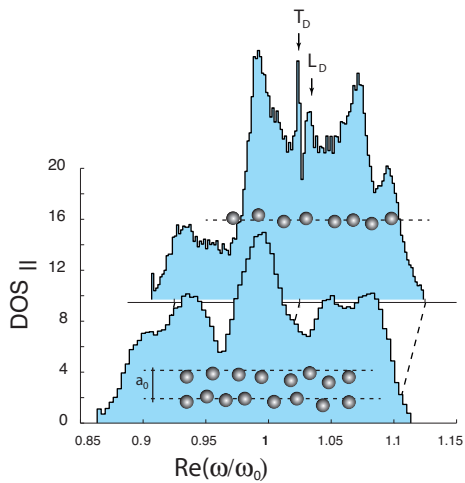}
\caption{Comparison of DOS$_{||}$ functions for double nanoparticle chain with $N=400$ and nanoparticle array $2\times N$ calculated in QS limit. The disorder parameters $\delta_{x}=0.15\ \ \delta_{y}=0.15$. The separation distance between the chain lines  is $a_{0}$.}
\label{2lineDOS}
\end{center}
\end{figure}

\section{Appendix}
\appendix 
\section{Calculation method}
\label{AppendixA}
  To  solve  the system (\ref{Gen2}) we imply  weak  dependence of the matrix {\bf H}  on frequency $\om$ and decompose it into a series around the frequency $\om=\om_{0}$:
$$
{\bf H}(\omega)={\bf H}(\omega_{0})+\delta {\bf H}(\omega_{0})\cdot (\omega-\omega_{0})+...={\bf H_{0}}+\delta {\bf H_{0}}\cdot (\omega-\omega_{0})+... 
$$
The matrix {\bf H$_{0}$} is matrix {\bf H}($\om$) calculated at the frequency $\om=\om_{0}$. 
The zero order eigen values can be computed from the system
\bee
\label{Gen4}
{\bf H_{0}}{ \bf d}=\lambda^{(0)} d.
\ene
The inverse polarizability is a zero order eigen value and the set of eigen frequencies are computed from relation $1/\alpha(\om_{i}^{(0)})=\lambda^{(0)}_{i}$, where $i$ is the number of eigen value and index $(0)$ denotes zero-order approximation. Then  we assume that $\lambda_{i}\approx\lambda^{(0)}_{i}+\lambda^{(1)}_{i}$ and the first order corrections $\lambda^{(1)}_{i}$ is computed with perturbation method
$$
\lambda^{(1)}_{i}=<\overline{d}_{i}^{0}|\delta{\bf H_{0}}\cdot (\omega_{i}-\omega_{0})|d_{i}^{0}>.
$$         
Here we use quasi-scalar product  $<\overline{a}|b>=\sum_{i}a_{i}b_{i}$ as matrix ${\bf H}$ is complex symmetric but non-Hermitian \cite{Markel2006}.

Such approach is valid as far as the perturbation term $\delta {\bf H_{0}}\cdot (\omega-\omega_{0})$ is small.  The matrix ${\bf H} (\om)$ depends on frequency in the form of $kr$, and $\delta{\bf H} (\om)\sim R/c{\bf H} (\om)$. The perturbation term is  smaller by factor of $\om_{0}R/c (\om_{i}-\om_{0})/\om_{0}=\kappa\delta \om_{i}$  than ${\bf H_{0}}$. From QS limit the estimation of band width gives $\delta \om \sim (R/a_{0})^{3}$. Thus, the perturbation method should be valid if  $\kappa(R/a_{0})^{3}\ll 1$. For the parameters of calculation used in this study $\kappa=0.15$ and $a_{0}=3R$ this condition is  fulfilled.

\section{Two line chain}
\label{AppendixB}

The  dimensionality  of the structure plays an important role for the observed DOS peak. To demonstrate that we have calculated the in-plane DOS function (see Section \ref{TLdrd}) for a two line chain of nanoparticles separated with distance $a_{0}$ from each other in $y$-direction. The calculated DOS for in-plane modes in QS limit is shown in fig. (\ref{2lineDOS}). One can  see that the fine structure of the DOS is totally smeared out for two line chain. Introduction of retardation  only enhances this effect. Thus, despite the fact that the system is embedded in 3D the quasi one-dimensional feature of nanoparticle chain is crucial for the observed Dyson peak.

\bibliography{References}

\end{document}